\input{aipcheck}

\documentclass[
    ,final            
  ]
  {aipproc}

\layoutstyle{8x11double}

\usepackage{color}
\begin{document}

\title{Flavor Data Constraints on the SUSY Parameter Space}


\author{Farvah Mahmoudi}{
  address={High-Energy Physics, Uppsala University, P.\,O.\,Box 535, 751\,21 Uppsala, Sweden},email=
{nazila.mahmoudi@tsl.uu.se}}

\begin{abstract}
We present an overview of the indirect constraints from flavor physics on supersymmetric models. During the past few years flavor data, and in particular $b \to s \gamma$ transitions, have been extensively used in order to constrain supersymmetric parameter spaces. We will briefly illustrate here the constraints obtained by a collection of low energy observables including FCNC transitions, rare decays, leptonic and semileptonic decays of $B$ mesons, as well as leptonic decays of $K$ mesons. The theoretical predictions can be obtained using the computer program SuperIso.
\end{abstract}

\maketitle


\section{Introduction}
Together with the direct searches for new particles and effects, indirect searches play an important role. Indeed, when supersymmetric particles appear as virtual states in Standard Model processes, they can reveal the indirect effects of supersymmetry. Therefore they can play a complementary role to the direct searches. 

In the constrained MSSM scenarios, such as CMSSM and NUHM which we investigate here, the fact that the number of free parameters is drastically reduced as compared to the general MSSM allows for the observables to probe deeply the structure of the considered scenario. The most commonly used indirect constraints along this line are the electroweak precision observables, the anomalous magnetic moment of the muon $(g-2)_\mu$, flavor physics observables and cosmological constraints from WMAP and the relic density. A quantification of the constraining power of each observable was presented in \cite{Allanach:arxiv:2008} in the large volume string scenario. Flavor physics observables, and in particular the isospin asymmetry in $B \to K^* \gamma$ decays are among the most important ones after the relic density and the muon anomalous magnetic moment. Some recent analyses studying constraints on the MSSM parameter space can be found in \cite{Carena:2006ai,Ellis:2007fu,Mahmoudi:2007gd,Heinemeyer:2008fb,Eriksson:2008cx}.

In the following, we present an overview of the most constraining flavor physics observables, namely the branching ratio of $B \to X_s \gamma$, the branching ratio of $B_s \to \mu^+ \mu^-$, the branching ratio of $B \to \tau \nu_\tau$, the branching ratio of $B \to D \tau \nu_\tau$ and the branching ratio of $K \to \mu \nu_\mu$. For each observable we determine the regions excluded in the NUHM parameter space. All the observables are calculated with the publicly available program SuperIso \cite{Mahmoudi:2007vz,Mahmoudi:2008tp}, and the spectra of SUSY particles are calculated using SOFTSUSY 2.0.18 \cite{Allanach:2001kg}. A brief description of each observable is provided below, and a more detailed description of the calculations can be found in \cite{Mahmoudi:2008tp}. 
For the CMSSM scenario, we provide the combined constraints by all observables in a single figure.

\section{Flavor observables}

\begin{figure}[!t]
\includegraphics[clip,bb=50 20 290 270,width=2.1cm]{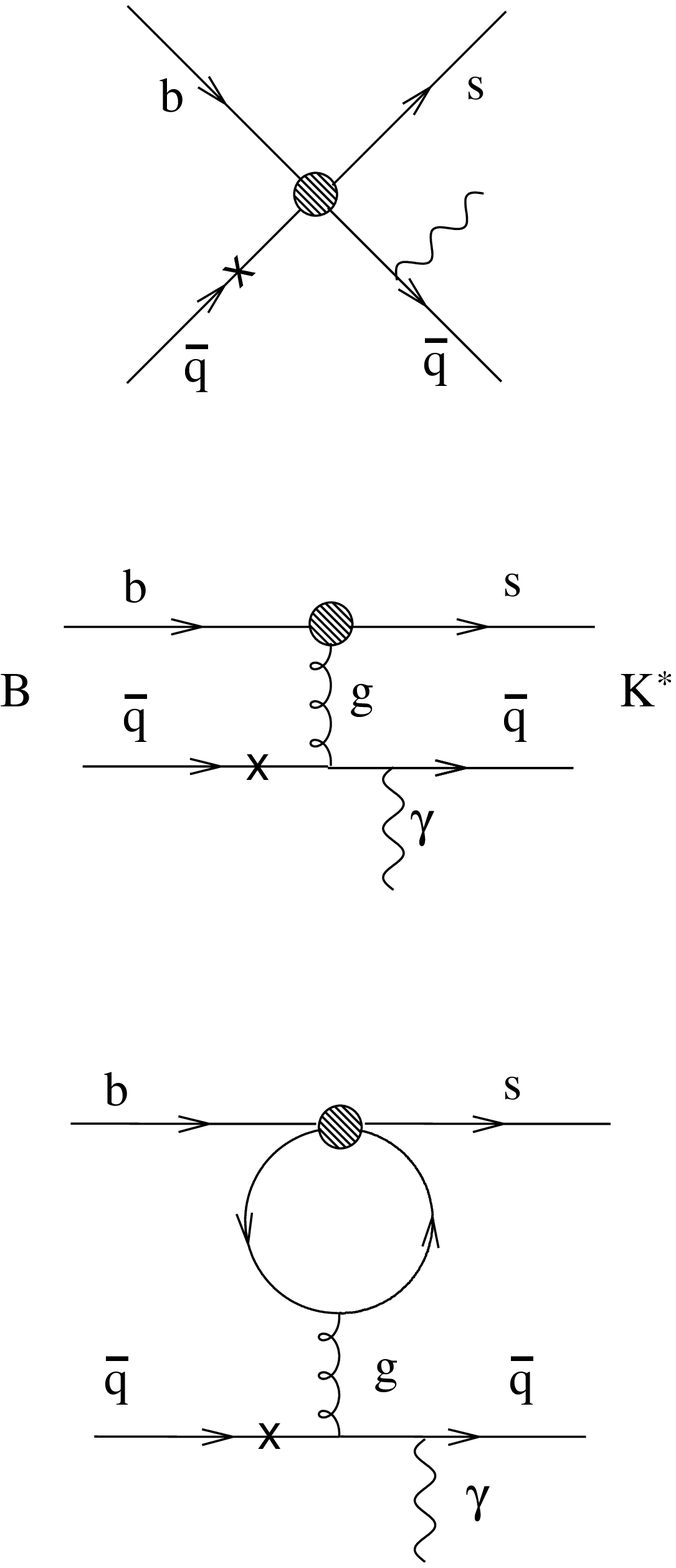}~~\includegraphics[clip,bb=50 560 240 720,width=2.2cm]{fig1}~~\includegraphics[clip,bb=25 330 280 495,width=2.9cm]{fig1}
\caption{Example of diagrams contributing to $B \to X_s \gamma$. \label{bsg1}}
\end{figure}
\begin{figure}[!t]
\includegraphics[width=2.1cm]{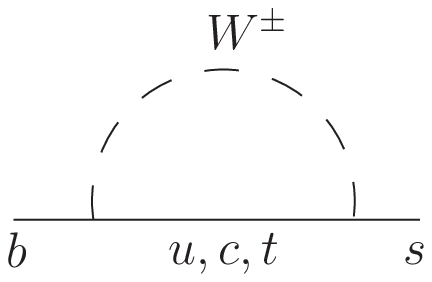} \hspace*{-0.4cm} \includegraphics[width=2.1cm]{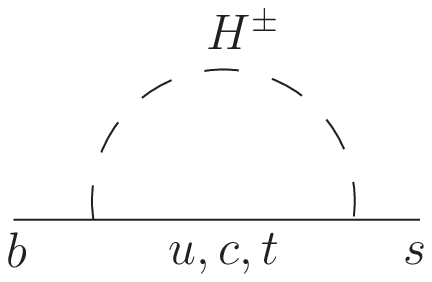} \hspace*{-0.4cm} 
\includegraphics[width=2.1cm]{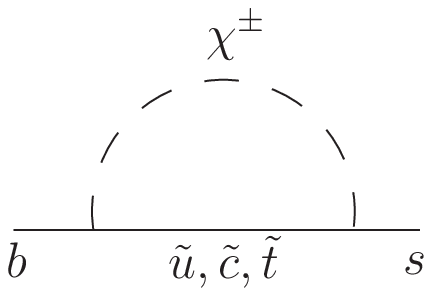} \hspace*{-0.4cm} \includegraphics[width=2.1cm]{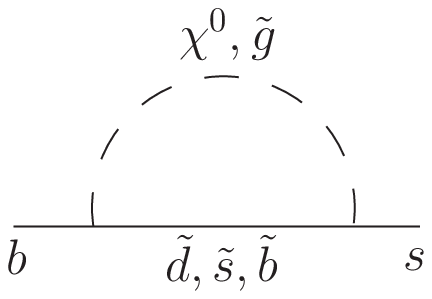}
\caption{Loops involved in $b\to s\gamma$ transitions.\label{bsg2}}
\end{figure}
Rare $B$ decays are very sensitive to new physics effects and can play an important role in disentangling different scenarios. 
The transition which is most often discussed in this context is the flavor changing neutral current process $b\to s\gamma$. Since this transition occurs first at one-loop level in the SM, the new physics contributions can be of comparable magnitude.
The penguin loops here involve $W$ boson in the Standard Model, and in addition loops from charged Higgs boson, chargino, neutralino and gluino for the MSSM as presented in Figs. \ref{bsg1} and \ref{bsg2}. The contribution of neutralino and gluino loops is negligible in minimal flavor violating scenarios.
Charged Higgs loop always adds constructively to the SM penguin. Thus, this observable is an effective tool to probe the 2HDM scenario. Chargino loops however can add constructively or destructively. If the interference is positive, it results in a great enhancement in the ${\rm BR}(b\to s\gamma)$, which become therefore a powerful observable. The latest combined experimental value for this branching ratio is reported by the Heavy Flavor Averaging Group (HFAG) \cite{Barberio:2008fa}:
\begin{equation}
{\rm BR}(\bar{B} \to X_s \gamma)_{{\rm exp}}=(3.52\pm0.23\pm0.09)\times10^{-4}.
\end{equation}
Combining the experimental result with the theoretical prediction, we find the range allowed at 95\% C.L.:
\begin{equation}
2.15\times 10^{-4}\leq {\rm BR}(\bar{B} \to X_s \gamma) \leq 4.89\times 10^{-4}.
\end{equation}
\begin{figure}[!t]
\includegraphics[width=6.1cm]{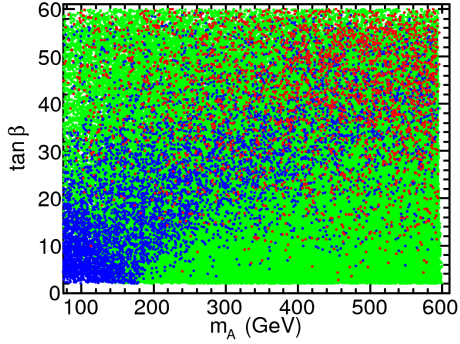}
\caption{Constraints on the NUHM parameter plane $(m_A,\tan\beta)$ from the branching ratio (blue) and isospin asymmetry (red) in $b\to s\gamma$ transitions. The excluded points are displayed in the foreground. \label{bsg3}}
\end{figure}

On the other hand, if the interference of the chargino contribution with the SM loop is destructive, the other interesting observable which opens up is the degree of isospin asymmetry in the exclusive decay of $B \to K^* \gamma$, defined as:
\begin{equation}
\Delta_{0\pm}\equiv\displaystyle\frac{\Gamma(\bar B^0\to\bar K^{*0}\gamma) - \Gamma(B^\pm \to K^{*\pm}\gamma)}{\Gamma(\bar B^0\to\bar K^{*0}\gamma) + \Gamma(B^\pm\to K^{*\pm}\gamma)}.
\end{equation} 

This observable often provides stricter limits on the parameters of different MSSM scenarios than the inclusive branching ratio, as has been shown earlier in \cite{Mahmoudi:2007gd}. 
Combining the most recent experimental values of BaBar \cite{Aubert:2008cy} and Belle \cite{Nakao:2004th}, including the experimental and theoretical uncertainties, the allowed range
\begin{equation}
-1.7\times 10^{-2} < \Delta_0 < 8.9\times 10^{-2}
\end{equation}
is obtained at 95\% C.L.

To investigate the NUHM parameter space, we perform scans over the parameters such that $m_0\in [50,2000]$, $m_{1/2}\in [50,2000]$,  $A_0\in [-2000,2000]$, $\mu \in [0,2000]$, $m_A \in [5,600]$ and $\tan\beta \in [1,60]$. Fig. \ref{bsg3} shows the resulting constraints in the $(m_A,\tan\beta)$ plane. The excluded points, in red from the isospin asymmetry and in blue from the branching ratio, are shown in the foreground. The remaining green points represent the allowed points. As can be seen in this figure, low values for $m_A$ and $\tan\beta$ are excluded by the branching ratio, whereas higher values of $\tan\beta$ are disfavored by the isospin asymmetry.

\begin{figure}
\includegraphics[width=3.1cm]{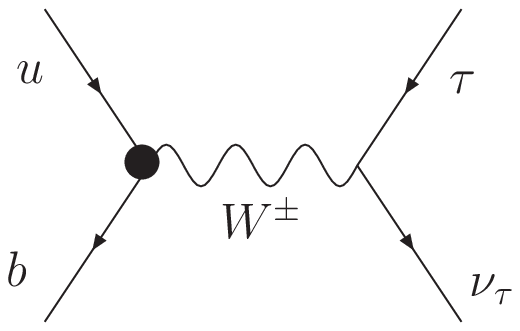} \hspace*{0.0cm} \includegraphics[width=3.1cm]{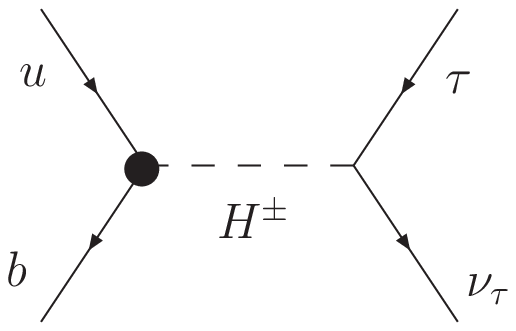} 
\caption{Diagrams contributing to $B_u \to \tau\nu_\tau$. \label{btn}}
\end{figure}

In contrast to the $b \to s\gamma$ transitions, the process $B_u\to\tau\nu_\tau$ can be mediated by a charged Higgs boson already at tree level in annihilation processes as shown in Fig.~\ref{btn}. Since this decay is helicity suppressed in the SM, whereas there is no such suppression for the charged Higgs exchange in the limit of high $\tan\beta$, these two contributions can be of similar magnitude \cite{Hou:1992sy}. This decay is thus very sensitive to charged Higgs boson and provides important constraints. The current HFAG value for ${\rm BR}(B_u \to \tau\nu_\tau)$ is \cite{Barberio:2008fa} 
\begin{equation}
{\rm BR}(B_u \to \tau\nu_\tau)_{\rm exp}=(1.41\pm 0.43)\times 10^{-4}.
\end{equation}

The new physics contribution from $H^+$ is expressed through the ratio of the MSSM over SM \cite{Akeroyd:2003zr}:

\begin{equation}
R^{\rm{MSSM}}_{\tau\nu_\tau}=\left[1-\left(\frac{m_B^2}{m_{H^+}^2}\right)\frac{\tan^2\beta}{1+\epsilon_0\tan\beta}\right]^2.
\label{eq:RMSSM}
\end{equation}
The leading SUSY-QCD corrections are included in this expression through $\epsilon_0$ \cite{Buras:2002vd}
\begin{equation}
\epsilon_0=-\frac{2\alpha_s}{3\pi}\frac{\mu}{m_{\tilde{g}}}H_2\left(\frac{m_Q^2}{m_{\tilde{g}}^2},\frac{m_D^2}{m_{\tilde{g}}^2}\right),
\label{eps0}
\end{equation}
where
\begin{equation}
H_2(x,y)=\frac{x\ln x}{(1-x)(x-y)}+\frac{y\ln y}{(1-y)(y-x)}.
\end{equation}

The calculation of ${\rm BR}(B_u \to \tau\nu_\tau)$ suffers however large uncertainties from the determination of $|V_{ub}|$, since different measurements of this quantity are incompatible.

\begin{figure}
\includegraphics[width=3.1cm]{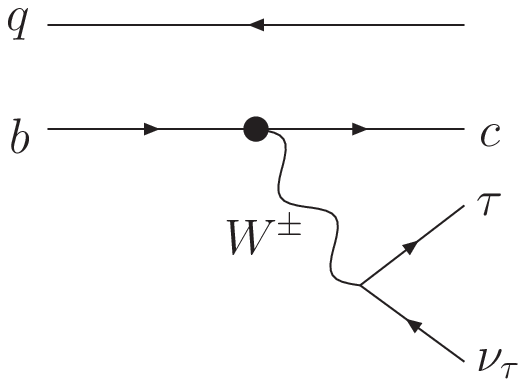} \hspace*{0.0cm} \includegraphics[width=3.1cm]{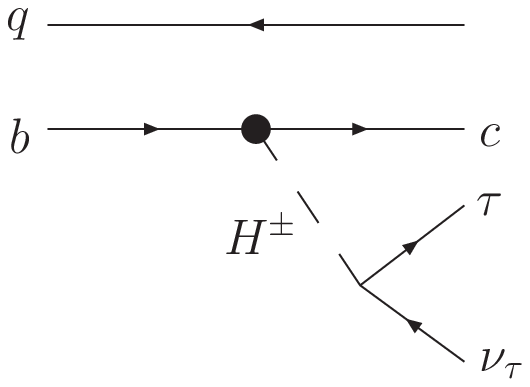} 
\caption{Diagrams contributing to $B\to D \tau \nu_\tau$. \label{bdtn}}
\end{figure}

The semileptonic decays $B \to D\ell\nu$, illustrated in  Fig.~\ref{bdtn}, have the advantage of depending on $|V_{cb}|$, which is known to better precision than $|V_{ub}|$. Due to the presence of at least two neutrinos in the final state, the experimental determination remains however very complex.
To reduce some of the theoretical uncertainties, we consider the following ratio \cite{Kamenik:2008tj}:
\begin{equation}
\xi_{D\ell\nu} \equiv \frac{{\rm BR}(B \to D \tau \nu_\tau)}{{\rm BR}(B \to D e \nu_e)},
\end{equation}

\begin{figure}[!t]
\includegraphics[width=6.1cm]{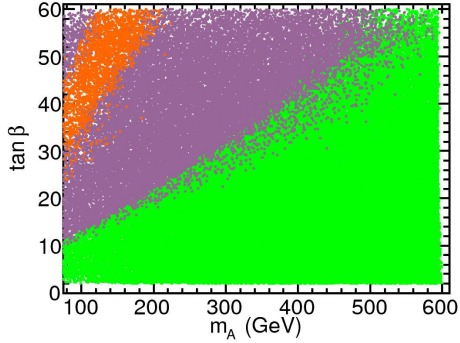}
\caption{Constraints on the NUHM parameter plane $(m_A,\tan\beta)$ from $B_u\to \tau \nu_\tau$ (purple) and $B\to D \tau \nu_\tau$ (orange). The excluded points are displayed in the foreground. \label{btn2}}
\end{figure}

Fig.~\ref{btn2} shows the effect in the $(m_A,\tan\beta)$ plane of imposing both the $R^{\rm{MSSM}}$ and $\xi_{D\ell\nu}$ constraints on the NUHM model points generated in the random grid, where the excluded points by ${\rm BR}(B_u\to \tau \nu_\tau)$ are shown in purple and by ${\rm BR}(B \to D \tau \nu_\tau)$ in orange. The excluded points are displayed in the foreground. The disfavored points by ${\rm BR}(B_u\to \tau \nu_\tau)$ fall in two disjoint regions with a gap filled out by ${\rm BR}(B \to D \tau \nu_\tau)$. To extract these results, the combined value of $|V_{ub}|$ obtained from the inclusive and exclusive semileptonic decays is used \cite{Amsler:2008zz}. The following allowed 95\% C.L. ranges for $R^{\rm{MSSM}}_{\tau\nu_\tau}$: 
\begin{equation}
0.53 < R^{\rm{MSSM}}_{\tau\nu_\tau} < 2.03,
\end{equation}
and for $\xi_{D\ell\nu}$:
\begin{equation}
15.1 \times 10^{-2} < \xi_{D\ell\nu} < 68.1 \times 10^{-2},
\end{equation}
are used. These two observables therefore play a complementary role, and improvements in the related experimental measurements would certainly be welcome.


The rare decay $B_s \to \mu^+ \mu^-$ proceeds via $Z^0$ penguin and box diagrams in the SM, and the branching ratio is therefore highly suppressed. This process has not been observed experimentally so far, and the experimental limit is an order of magnitude away from the SM prediction, allowing for substantial new physics contributions \cite{Bobeth:2001sq}.
The current experimental limit, derived by the CDF collaboration at 95\% C.L., is \cite{Aaltonen:2007kv}:
\begin{equation}
{\rm BR}(B_s \to \mu^+ \mu^-) < 5.8 \times 10^{-8}.
\end{equation}

\begin{figure}
\includegraphics[width=3.1cm]{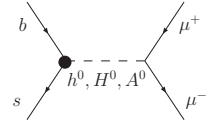} 
\caption{Diagrams contributing to $B_s \to \mu^+ \mu^-$ at high $\tan\beta$. \label{bmm}}
\end{figure}

In supersymmetry, for large values of $\tan\beta$, this decay can receive large contributions from neutral Higgs bosons 
illustrated in Fig.~\ref{bmm}.

Including theoretical uncertainties, we compare the MSSM prediction to the upper limit at 95\% C.L.
\begin{equation}
{\rm BR}(B_s \to \mu^+ \mu^-) < 6.6 \times 10^{-8},
\end{equation}
in order to explore the constraints obtained by this observable.

\begin{figure}[!t]
\includegraphics[width=6.1cm]{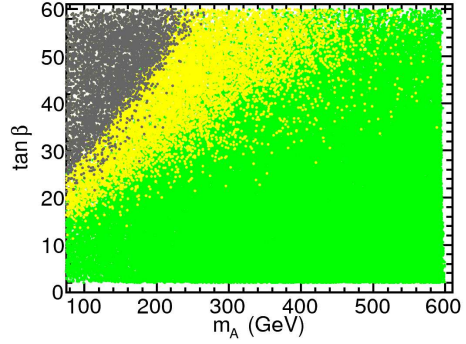}
\caption{Constraints on the NUHM parameter plane $(m_A,\tan\beta)$ from $B_s \to \mu^+ \mu^-$ (yellow) and $K \to \mu \nu_\mu$ (grey). The excluded points are displayed in the foreground. \label{bmm2}}
\end{figure}
%
\begin{figure}
\resizebox{.8\textwidth}{!}{
\includegraphics[width=6.1cm]{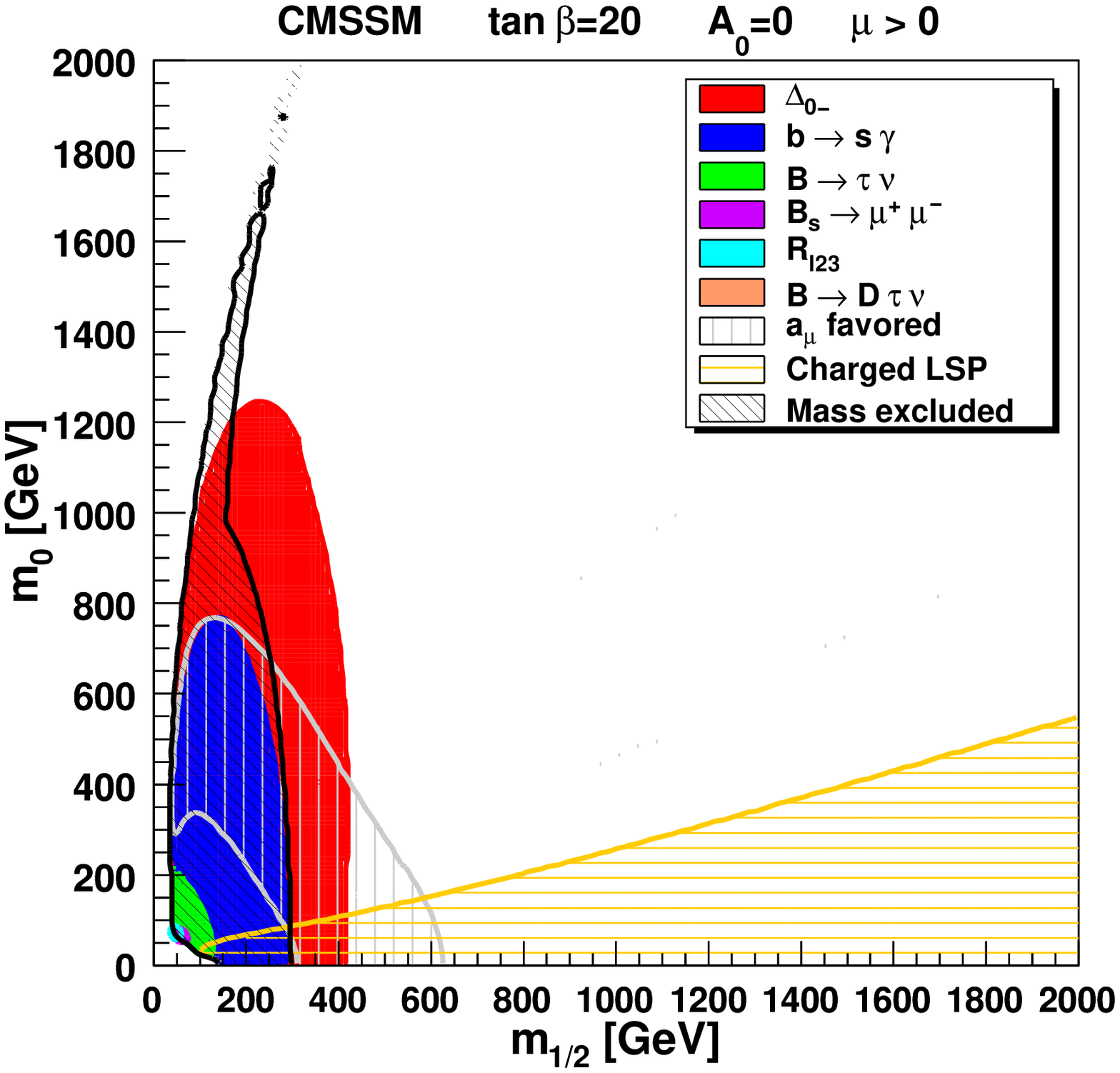}
\includegraphics[width=6.1cm]{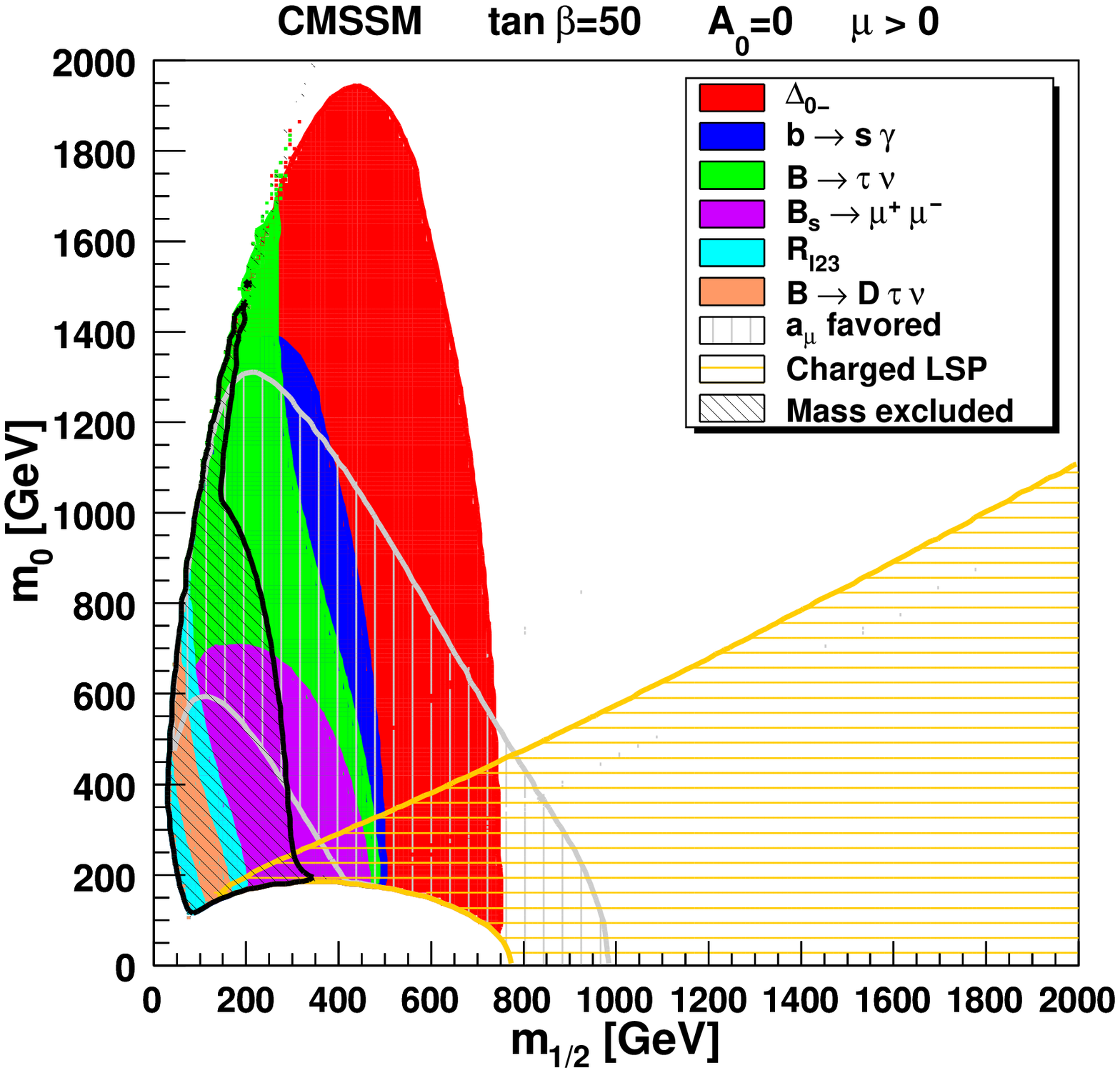}}
\caption{Constraints on the CMSSM parameter plane $(m_{1/2},m_0)$, for $A_0=0$ and $\tan\beta=20$ (left) and $\tan\beta=50$ (right). \label{cmssm}}
\end{figure}

Another decay process, which has many similarities to $B_u \to \tau \nu_\tau$, is $K \to \mu \nu_\mu$. This decay is also mediated via $W^+$ and $H^+$ at tree-level similarly to the diagrams in Fig. \ref{btn}. In the case of $K \to \mu \nu_\mu$ decays however the $H^+$ contribution is reduced by the coupling of $H^+$ to lighter quarks.
In order to decrease the theoretical uncertainties from $f_K$, the ratio of partial widths of ${\rm BR}(K \to \mu \nu_\mu)$ over ${\rm BR}(\pi \to \mu \nu_\mu)$ is usually considered. As suggested in \cite{Antonelli:2008jg}, we study the related quantity 
\begin{equation}
R_{\ell 23}=\left|1-\frac{m^2_{K^+}}{M^2_{H^+}}\left(1 - \frac{m_d}{m_s}\right)\frac{\tan^2\beta}{1+\epsilon_0\tan\beta}\right|,
\end{equation}
where $\ell i$ refers to leptonic decays with $i$ particles in the final state.
The MSSM prediction can be directly compared to the experimental value \cite{Antonelli:2008jg}
\begin{equation}
R_{\ell 23}=1.004\pm 0.007.
\label{Rl23}
\end{equation}
The resulting constraints from both $B_s\to\mu^+\mu^-$ and $K \to \mu \nu_\mu$ are shown in Fig.~\ref{bmm2}, where again the excluded points are displayed in the foreground.
The region which is almost completely excluded is for very small $m_{H^+}$ and large $\tan\beta$.
In the extraction of these results, the ratio $f_K/f_\pi$ has been fixed to the value $f_K/f_\pi=1.189\pm 0.007$ obtained from lattice QCD using staggered quarks \cite{Follana:2007uv} which acquires a reduced error bar. There are several methods to compute this ratio, and if other values are used instead, the ${\rm BR}(K \to \mu \nu_\mu)$ unfortunately provides no constraints on the studied NUHM parameters. 

Considering now the CMSSM scenario, we show in Fig.~\ref{cmssm} the combined results of the flavor observables in the $(m_{1/2}, m_0)$ plane for $A_0=0$, and $\tan\beta=20$ in the left figure and $\tan\beta=50$ in the right figure. For completeness, we present also the direct limits on the SUSY masses, the neutral LSP condition and the region favored by the muon anomalous magnetic moment in these figures. The flavor data constraints are more restrictive for higher values of $\tan\beta$ as can be seen in these figures.
\section{Conclusion}
Flavor data constraints appear among the most powerful indirect constraints as we have seen here. Phenomenologically, it is very important to investigate the results of the indirect constraints since they can provide us with important information before the LHC data, therefore they can play a complementary role to the direct searches on one hand, and can be very valuable for the consistency checks on the other hand when the data become available.
\begin{theacknowledgments}
I would like to thank the SUSY08 organizers, and in particular the SUSY Collider Phenomenology conveners for their invitation. Some of the results presented here have been achieved in collaboration with David Eriksson and Oscar St\aa l to whom I am grateful.
\end{theacknowledgments}



\bibliographystyle{aipproc}   

\bibliography{susy08_mahmoudi}


\end{document}